\documentclass[5p]{elsarticle}
\usepackage{hyperref}

\usepackage{url}

\usepackage{natbib}
\usepackage{relsize}

\usepackage{multirow}
\usepackage{arydshln}
\usepackage{amsmath}
\usepackage{float}

\newcommand{\mgcc}{\text{mg/cm}^3}

\newcommand{\db}{\text{dB}}
\newcommand{\BVTV}{\text{BV/TV}}
\newcommand{\BMD}{\text{BMD}}
\newcommand{\TMD}{\text{TMD}}
\newcommand{\SD}{\text{SD}}
\newcommand{\HS}{\text{H}}
\newcommand{\softplus}[1]{\text{plus}_{\epsilon}(#1)}
\newcommand{\fHS}{\text{H}_{\sigma}}
\newcommand{\NNBMD}{\text{NN}_{\text{BMD}}}
\newcommand{\NNSP}{\text{NN}_{\text{SP}}}
\newcommand{\mur}{\text{E}_{\text{r}}}
\newcommand{\mup}{\text{E}_{\text{p}}}
\newcommand{\varr}{\text{Var}_{\text{r}}}
\newcommand{\varp}{\text{Var}_{\text{p}}}

\journal{Computerized Medical Imaging and Graphics}

\begin{document}

\begin{frontmatter}

\title{Noise Reduction to Compute Tissue Mineral Density\\and Trabecular Bone Volume Fraction from Low Resolution QCT}


\author[uns,conicet]{Felix~Thomsen\corref{mycorrespondingauthor}}
\ead{felix.thomsen@uns.edu.ar}
\author[ceatic]{Jos\'e~M.~Fuertes Garc\'ia}
\author[ceatic]{Manuel~Lucena}
\author[uns]{Juan~Pisula}
\author[LCI]{Rodrigo~de~Luis~Garc\'{\i}a}
\author[Minden]{Jan~Borggrefe}
\author[uns,conicet]{Claudio~Delrieux}
\cortext[mycorrespondingauthor]{Corresponding author}
\address[uns]{National University of the South, DIEC, Bah\'ia Blanca, Argentina}
\address[conicet]{National Scientific and Technical Research Council, Buenos Aires, Argentina}
\address[ceatic]{University of Ja\'en, Ja\'en, Spain}
\address[LCI]{University of Valladolid, Valladolid, Spain}
\address[Minden]{Universit\"atsinstitut f\"ur Radiologie, Neuroradiologie und Nuklearmedizin, Minden, Germany}

\begin{abstract}
Micro-structural parameters of the thoracic or lumbar spine generally carry insufficient accuracy and precision for clinical \emph{in vivo} studies when assessed on quantitative computed tomography (QCT).
We propose a 3D convolutional neural network with specific loss functions for QCT noise reduction to compute micro-structural parameters such as tissue mineral density (TMD) and bone volume ratio (BV/TV) with significantly higher accuracy than using no or standard noise reduction filters.
The vertebra-phantom study contained high resolution peripheral and clinical CT scans with simulated \emph{in vivo} CT noise and nine repetitions of three different tube currents (100, 250 and 360~mAs).    
Five-fold cross validation was performed on $20466$ purely spongy pairs of noisy and ground-truth patches. 
Comparison of training and test errors revealed high robustness against over-fitting.
While not showing effects for the assessment of BMD and voxel-wise densities, the filter improved thoroughly the computation of TMD and BV/TV with respect to the unfiltered data.
Root-mean-square and accuracy errors of low resolution TMD and BV/TV decreased to less than $17\%$ of the initial values.
Furthermore filtered low resolution scans revealed still more TMD- and BV/TV-relevant information than high resolution CT scans, either unfiltered or filtered with two state-of-the-art standard denoising methods.	
The proposed architecture is threshold and rotational invariant, applicable on a wide range of image resolutions at once, and likely serves for an accurate computation of further micro-structural parameters.
Furthermore, it is less prone for over-fitting than neural networks that compute structural parameters directly.
In conclusion, the method is potentially important for the diagnosis of osteoporosis and other bone diseases since it allows to assess relevant 3D micro-structural information from standard low exposure CT protocols such as 100~mAs and 120~kVp.
\end{abstract}

\begin{keyword}
convolutional neural network \sep in vivo \sep local micro-structure \sep phantom study \sep regression
\end{keyword}

\end{frontmatter}


\section{Introduction}
Bone mineral density (BMD) explains around $70\%$ of bone stability under osteoporosis whereas the remaining $30\%$ of information is believed to be explained by qualitative micro-structural parameters~\cite{Thomsen2016LFD}.
BMD is invariant to computed tomography (CT) noise and stable over a large range of image resolutions. 
Qualitative micro-structural parameters like tissue mineral density (TMD), bone volume ratio (BV/TV), trabecular separation, thickness, or the structural model index are generally computed on a binary presentation of the volume. 
This segmentation into bone and marrow is obtained by applying a global threshold which is sensitive to CT noise and the low spatial resolution.
Hence, qualitative micro-structural parameters are biased and less stable under \emph{in vivo} conditions and are therefore labeled as ''apparent'' (e.g. app.TMD or app.BV/TV) and only accessible under clinical high-resolution quantitative computed tomography (HR-QCT) but not under standard- or low-resolution QCT (LR-QCT). 

There are two approaches to improve computation of existing micro-struc\-tu\-ral parameters under \emph{in vivo} conditions.
The first is to use novel algorithms that are implicitly robust against noise, as for instance done for the plate-to-rod ratio~\cite{Thomsen2016LFD}, trabecular number~\cite{Moreno2012estimation} or trabecular separation~\cite{Darabi2009Thickness}.
These methods, that are generally difficult to design, can be achieved by either reformulating the entire algorithm or by replacing the most fragile steps of the standard algorithms with more stable alternatives.
The second approach is based on machine learning, in particular convolutional neural networks (CNNs).
CNNs are widely used for segmentation, classification, denoising, and super-resolution applications~\cite{Shin2016Deep}, and can also be applied in regression problems.
Regression requires generally less training data than classification since models can be trained on small subsets of the data, and network architectures for regression require less refinement than those of classification problems.
Neural networks for classification are generally a combination of a convolutional and a fully connected neural network~\cite{Cintas2020}.
The most direct application of regression in CT imagery is noise removal or accentuation of high frequencies~\cite{You2018structurally}, thus enhancing the original CT volume to obtain a more suitable dataset to perform certain desired tasks with higher accuracy and precision.
Approaches of this kind were trained for instance with the root-mean-square error (RMSE) or the structural similarity index \cite{Wang2004Image} on a 3-layer CNN~\cite{Chen2017Low} or 7-layer auto-encoder U-Net~\cite{Chen2017Low2}. 
Those architectures obtained similar errors as the Wavelet based ''Block Matching and 3D (or 4D) Filtering Technique'' (BM3D or BM4D) when applied on images with a peak signal-to-noise ratio (PSNR) of around $30~\db$.
More advanced generative adversarial networks have been successfully used to transform volumes of tube current of $10$~mAs into volumes corresponding to a visually perceptible quality of $50$~mAs~\cite{Wolterink2017GAN}.
Combining two generative adversarial networks additionally allowed to solve the inverse problem: synthesize scanner deterioration and thus transform noise-free \emph{in silico} phantoms into simulated \emph{in situ} scans~\cite{Russ2019synthesis}. 

\begin{figure}
	\centering
	\includegraphics[width=\linewidth]{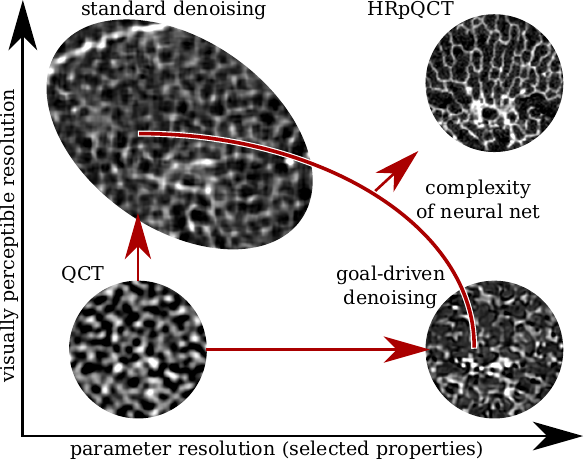}
	\caption{Standard denoising for visual perceptible resolution versus goal-driven denoising for particular properties only, for instance to compute BMD, TMD or BV/TV. Achieving simultaneously both requires much higher complexity and might generally be unattainable.}
	\label{fig:mapping}
\end{figure}
Our hypothesis is that specific structural parameters do not necessarily require a volume of general high visual detail.
In other words, visual quality {\em per se} is not of intrinsic diagnostic or clinical advantage in QCT.
Instead, we implement a goal-driven denoising straight to enhance computation of specific local micro-structural parameters (BMD, TMD and BV/TV, see Fig.~\ref{fig:mapping}).
There exist two approaches to enhance micro-structural computation with CNNs: First, an immediate estimation of local micro-structural parameters with deep-learning, and second, a goal-driven denoising for the subsequent assessment of specific micro-structural parameters with conventional (non-CNN) methods.

We employed here the second approach, a low-level volume filter which transforms the input volume into one that allows micro-structural assessment with higher accuracy and precision, thereby neither implementing explicit visual noise reduction nor computing implicitly the micro-structural parameters.
This approach allows the application of different micro-structural parameters, thresholds, and different noise levels with the same neural network and in the same moment.
It requires however the acquisition of specific concurrently applicable smooth and local loss functions for each micro-structural parameter. 
In this prototype study, we chose TMD and BV/TV for several reasons as parameters under investigation.
They 1)~are commonly known and widely used, 2)~contain simple concurrent implementations, 3)~are not yet accurately accessible on LR-QCT, and 4)~are of high importance on \emph{in vivo} studies to analyze the treatment of osteoporosis~\cite{Gluer2013Comparative}, to quantify multiple myeloma~\cite{Borggrefe2015Olymp} or to predict bone failure load~\cite{Lu2015role,Thomsen2016LFD}.
We derive a sensible architecture of the neural network and report the used hyper-parameters of network architecture (learning rate, batch size, etc.) allowing to repeat the analysis and to apply the method on further micro-structural parameters.

\section{Materials and Methods}
In this section, we describe the sampling and scanning procedure of the data (\ref{MM:Samples}), the architecture of the CNN and its novel loss functions (\ref{MM:CNN}), and the experiments and applied statistics (\ref{MM:Stats}). 
We used Python~(v3.6) with SimpleITK~(v1.2.3) to import and register the DICOM volumes, and PyTorch~(v1.2) for the design and application of the neural network.
We trained the network on an NVidia GP102 Titan XP graphics processing unit.

\subsection{Sampling of patches}\label{MM:Samples}
\begin{figure}[ht!]
	\centering
	\includegraphics[width=\linewidth]{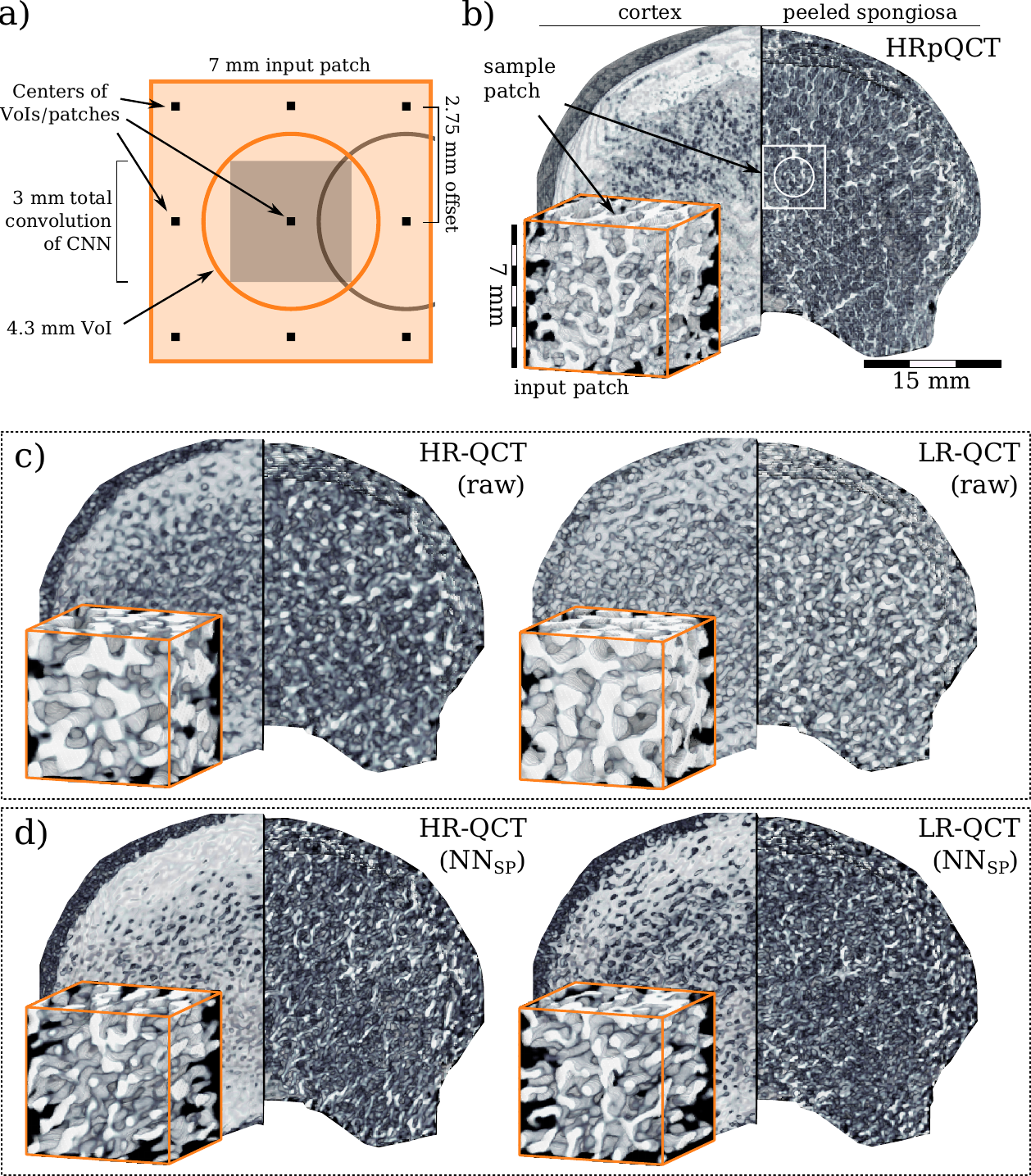}
	\caption{a) Size and shape of input patch, VoI, offset and total convolution size of the CNN. b) Ray-casting of a HRpQCT scan, the peeled spongiosa (right) was used to extract patches (front left). c) Same vertebra under HR- and LR-QCT without and d) after filtering with $\NNSP$.}
	\label{fig:patching}
\end{figure}
Twelve human vertebrae (T12 and L1) have been embedded into epoxy resin without damaging any trabeculae to become cylindrical vertebra phantoms that fit perfectly into an abdomen phantom (Model 235, Computerized Imaging Reference Systems Inc, Norfolk, VA, USA) allowing to simulate \emph{in vivo} noise.
The vertebra phantoms have been scanned on a high-resolution peripheral QCT (HR\-pQCT) with isotropic resolution of $82$ $\mu\text{m}$, $59.4$~kVp and $900~\mu$As (Xtre\-me\-CT I, Scanco Medical AG, Br\"utti\-sel\-len, Switzerland) and automatically calibrated to density values.
The spongiosa has been peeled from the cortex with a semi-automatic procedure~\cite{Thomsen2016SI} and down-sampled to $172~\mu\text{m}  \times 172~\mu\text{m} \times 340~\mu\text{m}$.
For the purpose of reducing the cost and complexity of the main experiment, we analyzed size, BMD, BV/TV and TMD of all segmented vertebrae and identified a subset of five prototypes that contained the same information as the entire set of vertebrae, see also discussion section.

For the main experiment we scanned the subset of five vertebra phantoms with the abdomen phantom on a clinical CT scanner (iCT 256, Philips, Amsterdam, Netherlands). 
We used three different noise levels by scanning each phantom with low- ($100$~mAs, LR-QCT), mid-term- ($250$~mAs, MR-QCT) and high-resolution tube currents ($360$~mAs, HR-QCT), each with three repetitions and without replacement to analyze simulated \emph{in vivo} noise independently of any positioning-artifacts.
Other parameters have been set to $120$~kVp, pitch 1, slice thickness $0.67$~mm, standard bone kernel (YB) and field of view $88$~mm, corresponding to a resolution of $172~\mu\text{m}\times172~\mu\text{m}\times340~\mu\text{m}$.
A detailed description and figures of the setup but with a different scanner can be found elsewhere~\cite{Thomsen2016LFD,Thomsen2017PHD}.  

All simulated \emph{in vivo} QCT reconstructions were separately registered to the corresponding HRpQCT volumes with a rigid 5-step-pyramid approach and B-Spline interpolation.
We calibrated each QCT volume separately with the per-voxel correlation between low-pass maps (Eq.~\ref{Eq:BMD} with $d = 5\text{mm}$) of QCT and calibrated HRpQCT volumes.
We extracted $2274$ purely spongy patches from the five phantoms with size $41 \times 41 \times 21$ voxels (isotropic box with diameter $7$ mm) and regular offset $16 \times 16 \times 8$ voxels ($2.752$ mm) yielding with three repetitions for each of the three tube currents a total of $20466$ different HRpQCT-QCT-pairs of patches for training and testing.
Figure~\ref{fig:patching} shows sample projections of the HRpQCT, HR-QCT and LR-QCT scans before and after applying the presented filter method $\NNSP$.

\subsection{Design of the convolutional neural network}\label{MM:CNN}
\begin{figure*}
	\centering
	\includegraphics[width=\linewidth]{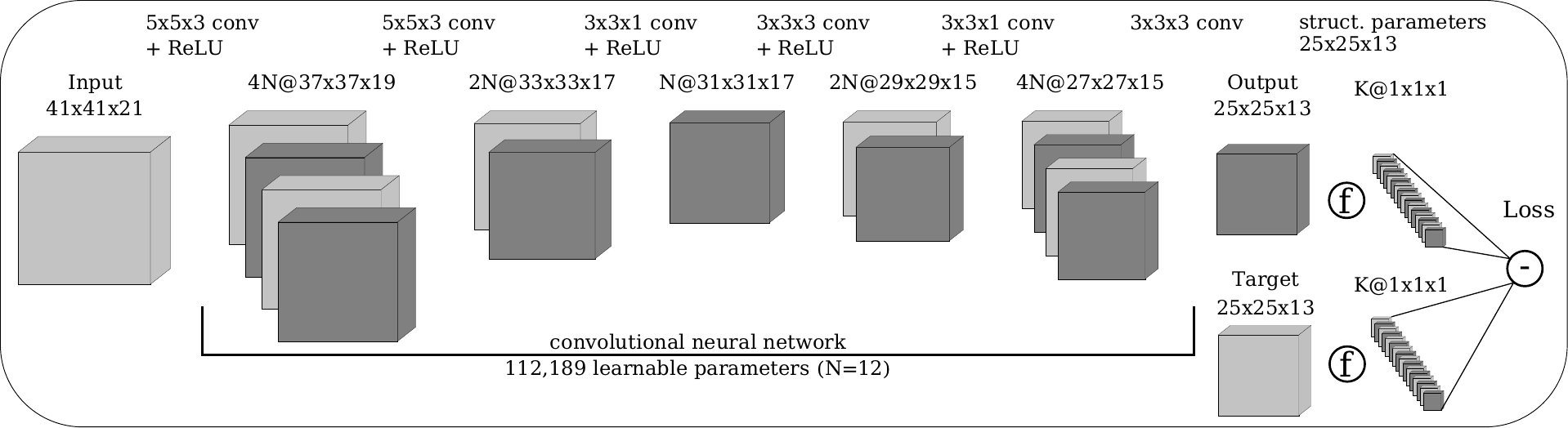}
	\caption{CNN with total mask size of $17\times17\times9$ voxels. The loss for back-propagation is computed on a set of $K$ structural parameters that are obtained outside the CNN on a neighborhood of $25 \times 25 \times 13$ voxels (isotropic box of $4.3$ mm).}
	\label{fig:nets}
\end{figure*} 	
The CNN implements a denoising filter for the subsequent computation of structural parameters.
It is defined on a local neighborhood of $17\times17\times9$ voxels, hence the neural network consumes the border of the input patch and remains only the center of $25\times25\times13$ voxels (isotropic box of $4.3$ mm) for the application of structural parameters.   
The number of convolutions per layer decreases in the first half and increases again in the second with in total six convolution layers: two $5\times5\times3$ layers followed by four alternating $3\times3\times1$ and $3\times3\times3$ layers, all but the last layer are followed by rectified linear units (ReLU), Fig.~\ref{fig:nets}.
Omitting the activation function of the last layer is required for regression problems~\cite{Chen2017Low2} and not using padding is sensible when training local filters~\cite{Wolterink2017GAN}; for the application to an entire vertebra in production mode we enabled padding again.   
Application of this architecture to an entire volume in production mode took less than 2 seconds.
Samples of the input- and output-patches are shown in Fig.~\ref{fig:patching}.

We implemented local structural parameters that resemble global parameters: bone mineral density ($\BMD$), bone volume ratio ($\BVTV$), and tissue mineral density ($\TMD$).
$\BMD$ is the average density of the calibrated input volume $V$.
$\BVTV_t$ is the ratio of bone to total volume when segmenting with threshold~$t$ and $\TMD_t$ is the density of segmented bone. 
The standard (global) formulas read
\begin{alignat}{2}
&\BMD(V)&=& \; \frac{1}{n} \sum_{\vec{x}} V(\vec{x}),\\
&\BVTV_t(V)&=& \; \frac{1}{n} \sum_{\vec{x}} \HS(V(\vec{x})-t), \text{and}\\
&\TMD_t(V)&=& \; \frac{\sum_{\vec{x}} \HS(V(\vec{x})-t) \; V(\vec{x})}{\sum_{\vec{x}} \HS(V(\vec{x})-t)},
\end{alignat}
with $\HS(x)$ the Heaviside function, $\vec{x}$ a voxel index of the global volume of interest (VoI) and $n$ the number of voxels of the VoI.

For the local parameters we needed to derive new formulas that 1)~allow computation on a very restricted local VoI instead of a global VoI per-vertebra, and 2)~are differentiable to be applicable for back-propagation.
Local and weighted parameters have been computed from spherical VoIs. 
Those VoIs were defined with convolution masks $\mathbf{N}_d$ with diameter $d$ and constant weight inside the enclosed sphere.
Spherical VoIs are superior to box-shaped VoIs regarding shape-compactness (e.g unit surface-area-to-volume ratio) and rotational invariance, thus micro-structural parameters of spherical VoIs are robust and independent of the alignment of the vertebrae~\cite{Ketcham2004,Thomsen2017PHD}.

Smoothness was achieved by defining parameters that are strictly monotone in~$t$. 
We defined a)~the softplus function $\softplus{x}$ that is a smooth version of $\max\{0,x\}$ parameterized with a fuzziness factor $\epsilon$ and with $\softplus{x}\approx x$ for $x>3.5\,\epsilon$:
\begin{equation}
\softplus{x} = \epsilon \ln\left(1+\exp\left(\frac{x}{\epsilon}-1\right)\right)+\epsilon,
\label{Eq:SoftMax}
\end{equation}
and b)~a fuzzy binarization $\fHS(x-t)$~\cite{Krebs2009Fuzzy}, a sigmoid version of the Heaviside function $\HS(x-t)$ with $\sigma>0$:
\begin{equation}
\fHS(x-t) =\left(1+\exp\left(\frac{t-x}{\sigma}\right)\right)^{-1}.
\label{Eq:fHS}
\end{equation}
Local weighted texture maps read then
\begin{alignat}{2}
&\BMD_d(V)&=&\left(V \ast \mathbf{N}_d\right)\label{Eq:BMD}\\
&\BVTV_{d,t}(V)&=&\left(\fHS(V-t) \ast \mathbf{N}_d\right),\\
&\TMD_{d,t}(V)&=&\left(\frac{(V \cdot \fHS(V-t)) \ast \mathbf{N}_d}{\softplus{\fHS(V-t)\ast \mathbf{N}_d}}\right), \label{Eq:locTMD}
\end{alignat}	
with $\ast$ the convolution operator, $\cdot$ the point-wise multiplication, $\epsilon=10^{-4}$ a small number and $\sigma=10~\mgcc$ a fuzzy scale.
Figure~\ref{fig:filterResults}~(HRpQCT) shows samples of those texture maps with $t=225~\mgcc$ and $d=4.3$~mm.
We additionally employed the local standard deviation~\cite{Thomsen2016MRI}:
\begin{equation}
\SD_d(V) = \sqrt{\frac{\left(V^2 \ast \mathbf{N}_d\right)-\left(V \ast \mathbf{N}_d\right)^2}{1-\sum_{\vec{x}}\mathbf{N}_d^2(\vec{x})}}.\\
\end{equation} 
An elementary loss function $L_f$ based on the mean-square error (MSE) was defined for each structural map.
For threshold independent parameters $f \in \{\BMD_d,\SD_d\}$, it reads: 
\begin{equation}
L_f = a_f \, \text{MSE}(f(x),f(y))\label{Eq:Loss1}
\end{equation}  
with $x$ the input and $y$ the ground truth data, replicated to gain the same number of repetitions as $x$ and $a_f$ a normalization factor to scale $L_f$ in average to one when computed over unfiltered data.

Threshold dependent parameters have been treated differently. 
The choice of the threshold depends on various factors such as the study design, image quality, patient's size and condition. 
Common thresholds vary between $200$ and $250~\mgcc$ and one might either apply a fixed threshold for an entire study, requiring some a-priori analysis~\cite{Borggrefe2015Olymp}, or select individual thresholds depending on the noise profile or by fixing BV/TV~\cite{Thomsen2016LFD}.
To gain threshold independence for $f_t \in \{\TMD_{d,t},\BVTV_{d,t}\}$ we designed a loss on multiple thresholds $t \in \{125,150,\cdots,325\}~\mgcc$:
\begin{equation}
L_f = a_f \, \sum_t b_t \, \text{MSE}(f_t(x),f_t(y))\label{Eq:MTLoss}
\end{equation}
with $x$, $y$ and $a_f$ as chosen as in Eq. \ref{Eq:Loss1}. The factor $b_t$ is sampled from the normal distribution with $\sigma=100~\mgcc$ and $\mu=225~\mgcc$ and normalized to $\sum_t b_t =1$ and serves as an ad-hoc fuzzy range of the most reasonable thresholds.
The compound loss is the weighted sum of individual losses $\{L_{f_1},\ldots,L_{f_k}\}$
\begin{equation}	
L = \sum_f w_f \, L_f  \label{Eq:compoundLoss}
\end{equation}
with $w_f$ the contribution of $L_f$ to the entire loss.

Back-propagation of the neural network was managed with the Adaptive Moment Estimation optimizer~\cite{Kingma2014adam} with an initial learning rate of $10^{-3.5}$ and default adjustment rates of $\beta_1=0.9$ and $\beta_2=1-10^{-3}$. 
During training and for each adjustment of the neural network, the compound loss has been computed over $64$ pairs of HRpQCT and filtered QCT patches, randomly taken from $16$ HRpQCT patches each with four of the nine available LR-, MR- or HR-QCT patches. 

The $20466$ pairs of patches have been divided into $70\%$ training ($14328$ pairs), $10\%$ validation ($2043$ pairs) and $20\%$ test set ($4095$ pairs with $1365$ for each of the three tube currents) and by not sharing any of the $2274$ different physical coordinates between cohorts.
We augmented the training data by factor $16$ to $229248$ pairs by applying all combinations of axial symmetric rotations and reflections that resulted in the same patch size, thereby encouraging rotational invariant filter solutions.
Validation and testing have been performed without data-augmentation.
Training was aborted when the loss, computed on the validation set, did not improve for $20$ epochs, yielding a total execution time of $6-8$ hours and about 140 epochs per fold.
We repeated this procedure five times with disjoint training, validation and test-sets to employ a cross-validation analysis with fixed hyper-parameters.

\subsection{Experiments and statistical analyses}\label{MM:Stats}
Three different denoising filters have been used: 
1)~We trained a neural network $\NNSP$ specifically to achieve a goal-driven denoising for the accurate computation of TMD and BV/TV independently of the used tube current and chosen threshold.
This has been done with the multi-threshold loss (Eq.\ref{Eq:MTLoss}), and by using concurrently patches of all three tube currents.
To maintain accuracy of the BMD and to support generalization of the noise reduction filter, we included also a loss of the BMD.
For $\NNSP$ we used VoIs covering the entire output-patches with $d=4.3$~mm (bounding box of $25\times25\times13$ voxels). 
Weights $w_f$ were $0.64$ for TMD, $0.32$ for BV/TV and $0.04$ for BMD, see Eq.\ref{Eq:compoundLoss}, thereby optimizing twice as much TMD than BV/TV, and dedicating only little to BMD.

2)~A state-of-the-art 3D noise reduction algorithm, the ''Block Matching and 4D Filtering'' (BM4D), has been deployed.
BM4D was reported to perform equally to a 3-layer CNN for CT noise reduction~\cite{Chen2017Low} and only slightly worse than a 7-layer U-Net~\cite{Chen2017Low2}, but both for data with less noise.   
The implementation was obtained from the Tampere University of Technology (\url{http://www.cs.tut.fi/~foi/GCF-BM3D/}).
We used automatic Gaussian noise estimation, normal noise profile and disabled Wiener filtering.

3)~Since BM4D is designed for higher SNRs we alternatively trained the reported neural network architecture with loss functions for pure denoising ($\NNBMD$), thereby using losses on BMD and SD but not on TMD and BV/TV.
The combination of losses on BMD and SD resembles in parts the structural similarity loss which is sometimes used for CT denoising~\cite{You2018structurally}.  
We used the voxel-wise MSE and VoIs with bounding boxes of $3\times3\times1$~voxels ($d_1$), $9\times9\times5$~voxels ($d_2$) and $17\times17\times9$~voxels ($d_3$).
Weights for the compound loss were $w=0.4$ for the voxel-wise MSE, $w=0.1$ for the BMD-loss with $d_1$ and $d_2$, and $w=0.4$ for the SD-loss with $d_3$, thereby focusing in particular on the voxel-wise error and penalizing over-smoothing with the SD-loss.

We computed statistics per-voxel and on the local maps of BMD, TMD and BV/TV of the raw and filtered data.
Standard statistics were the difference of averages with HRpQCT ($\Delta$Avg), the standard deviation (SD), root-mean-square error (RMSE) and adjusted R$^2$.
The peak-signal-to-noise ratio was reported as:
\begin{equation}
\text{PSNR} = 20 [\log(3\;\text{SD}(y))-\log(\text{RMSE}(x,y))]
\end{equation}
with $x$ the filtered or input data, $y$ the ground truth data, $\text{RMSE}(x,y)$ the root-mean-square error and $\text{SD}(y)$ the standard deviation on $y$.
The maximum possible range of intensities, part of the common definition of PSNR, is a rather arbitrary value for the considered specific transformations, thus we replaced it with $3\;\text{SD}(y)$.

Since our experiment contained repetitions, we were able to derive accuracy and precision metrics.
We extended standard metrics~\cite{Gluer1995Precision} for the application in machine learning applications, in particular with small number of repetitions. 
The accuracy error (AE) is a measure of the average bias and the precision error (PE) is the uncertainty of a particular instance.    
More precisely, AE is the RMSE between the average repetition of the noisy data and ground-truth, but corrected for the statistical uncertainty of the average operator. 
PE is the square root of the mean variance of repetitions normalized to the ground-truth scale, hence corrected for small standard deviations due to down-scaling. 
Both errors are expressed in the same unit and scale as the ground-truth data:
\begin{alignat}{2}
&\text{PE} &=& \sqrt{\frac{\mup(\varr(x)) \; \varp(y)}{\mur(\varp(x))}} \text{,}\\
&\text{AE} &=& \sqrt{\max\left\{0,\text{MSE}(\mur(x),y)-\frac{\mup(\varr(x))}{N-1}\right\}}
\end{alignat}        
with $x$ the input, $y$ the ground truth data, $\mup$ and $\varp$ the arithmetic mean and sample variance over all data points of the same repetition, $\mur$ and $\varr$ the arithmetic mean and sample variance over $N=3$ repetitions per tube current (or $N=9$ repetitions for all tube currents) and $\text{MSE}(x,y)$ the mean square error.     
We reported the mean and half-range for each statistic, computed on a 5-fold cross validation of the neural networks.
Statistics of the raw- and BM4D-filtered data have been computed on the same 5 sets used for testing during cross-validation.   

\section{Results}
\begin{figure*}
	\centering
	\includegraphics[width=\linewidth]{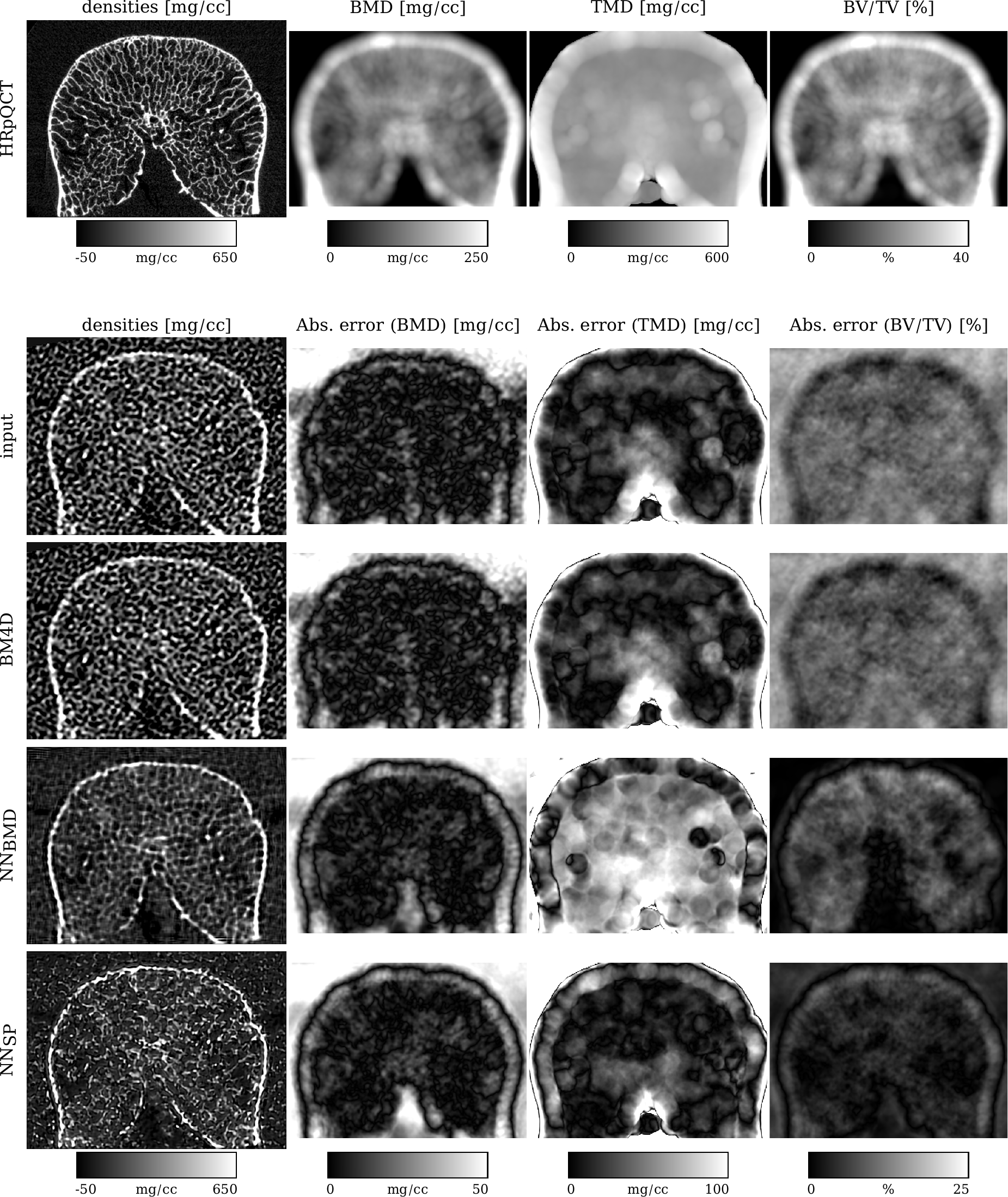}
	\caption{First row: HRpQCT ground truth input and structural maps of BMD, TMD and BV/TV with $t=225~\mgcc$ and a neighborhood size of $25\times25\times12$ voxels (4.3mm).
		Second to fifth row: Sample outputs of the unfiltered, BM4D-filtered and NN-filtered MR-QCT data (250~mAs) and absolute errors with HRpQCT of the structural maps BMD, TMD and BV/TV, dark regions are lowest errors.}
	\label{fig:filterResults}
\end{figure*}
Figure~\ref{fig:filterResults} shows sample outputs of the unfiltered and filtered data.
Structural maps are restricted to $4.3$~mm VoIs which fit entirely in the input volume, thus the structural maps are $24\times24\times12$ voxels smaller than the input or filtered volumes (that were generated with padding). 
Effects of BM4D are visually nearly undetectable.
$\NNSP$ obtained lowest errors of TMD and BV/TV. 
Both neural networks performed better on the spongiosa than on the cortex and sub-cortex which were excluded during training.

\begin{table*}
	\caption{Statistics of HR-QCT data: mean $\pm$ half range of test sets of 5-fold cross-validation. In bold (except for SD): best values per setting.}
	\centering
	\resizebox{\linewidth}{!}
	{\begin{tabular}{ll|ccccccc}
			Parameter&Filter&$\Delta$ Avg &SD &RMSE&AE&PE&PSNR&adj.R$^2$\\ 
			\hline\hline
			\multirow{4}{60pt}{per voxel [$\mgcc$]}
			&$-$&$1.61 \pm 0.64$&$191.36 \pm 0.36$&$199.43 \pm 0.54$&$\mathbf{71.24} \pm 1.36$&$114.50 \pm 0.79$&$12.95 \pm 0.14$&$7.14 \% \pm 0.32\%$\\
			&BM4D&$1.61 \pm 0.61$&$182.59 \pm 0.35$&$191.45 \pm 0.55$&$75.25 \pm 1.23$&$113.41 \pm 0.77$&$13.77 \pm 0.14$&$7.65 \% \pm 0.34\%$\\
			&$\NNBMD$&$0.51 \pm 1.04$&$101.01 \pm 1.06$&$\mathbf{130.42} \pm 1.01$&$98.38 \pm 0.93$&$\mathbf{99.69} \pm 1.06$&$\mathbf{21.45} \pm 0.14$&$\mathbf{13.23\%} \pm 0.42\%$\\
			&$\NNSP$&$\mathbf{0.04} \pm 0.60$&$130.96 \pm 2.33$&$157.76 \pm 2.42$&$95.06 \pm 1.40$&$113.09 \pm 2.29$&$17.64 \pm 0.19$&$6.40 \% \pm 1.12\%$\\
			
			\hdashline
			\multirow{4}{50pt}{BMD [$\mgcc$]}
			&$-$&$1.49 \pm 0.32$&$34.12 \pm 0.56$&$8.31 \pm 0.19$&$6.38 \pm 0.27$&$4.96 \pm 0.16$&$50.34 \pm 0.79$&$94.37 \% \pm 0.48\%$\\
			&BM4D&$1.49 \pm 0.32$&$34.11 \pm 0.57$&$8.28 \pm 0.20$&$6.39 \pm 0.27$&$4.91 \pm 0.16$&$50.42 \pm 0.81$&$94.41 \% \pm 0.48\%$\\ 
			&$\NNBMD$&$0.15 \pm 1.00$&$33.78 \pm 0.96$&$\mathbf{6.71} \pm 0.24$&$\mathbf{5.76} \pm 0.30$&$\mathbf{3.23} \pm 0.16$&$\mathbf{54.65} \pm 1.03$&$\mathbf{96.24\%} \pm 0.37\%$\\
			&$\NNSP$&$\mathbf{-0.12} \pm 1.06$&$33.74 \pm 1.02$&$6.85 \pm 0.26$&$5.78 \pm 0.29$&$3.46 \pm 0.27$&$54.21 \pm 0.75$&$96.10 \% \pm 0.24\%$\\
			
			\hdashline
			\multirow{4}{50pt}{TMD [$\mgcc$]}
			&$-$&$5.37 \pm 0.63$&$19.30 \pm 0.50$&$28.82 \pm 0.36$&$27.48 \pm 0.36$&$12.43 \pm 0.25$&$22.66 \pm 0.61$&$15.90 \% \pm 3.34\%$\\
			&BM4D&$-1.57 \pm 0.63$&$18.85 \pm 0.55$&$27.72 \pm 0.38$&$26.34 \pm 0.36$&$12.66 \pm 0.28$&$23.44 \pm 0.66$&$18.03 \% \pm 3.53\%$\\
			&$\NNBMD$&$-44.94 \pm 2.61$&$18.62 \pm 3.14$&$49.71 \pm 2.24$&$49.23 \pm 2.28$&$\mathbf{10.28} \pm 1.55$&$11.77 \pm 1.44$&$50.14 \% \pm 2.80\%$\\
			&$\NNSP$&$\mathbf{-0.05} \pm 3.75$&$23.90 \pm 0.73$&$\mathbf{17.70} \pm 1.10$&$\mathbf{14.35} \pm 1.32$&$11.97 \pm 0.71$&$\mathbf{32.42} \pm 1.19$&$\mathbf{65.30 \%} \pm 4.66\%$\\
			
			\hdashline
			\multirow{4}{50pt}{BV/TV [$\%$]}
			&$-$&$11.02 \pm 0.09$&$6.58 \pm 0.18$&$11.51 \pm 0.08$&$11.41 \pm 0.07$&$1.48 \pm 0.06$&$11.39 \pm 0.67$&$76.88 \% \pm 1.03\%$\\
			&BM4D&$10.10 \pm 0.08$&$6.81 \pm 0.19$&$10.63 \pm 0.08$&$10.51 \pm 0.08$&$1.48 \pm 0.06$&$12.98 \pm 0.65$&$77.47 \% \pm 1.00\%$\\ 
			&$\NNBMD$&$-3.39 \pm 0.35$&$7.56 \pm 0.19$&$4.23 \pm 0.36$&$4.03 \pm 0.39$&$\mathbf{1.05} \pm 0.06$&$31.47 \pm 2.00$&$89.07 \% \pm 0.93\%$\\
			&$\NNSP$&$\mathbf{0.38} \pm 0.38$&$6.70 \pm 0.15$&$\mathbf{2.16} \pm 0.12$&$\mathbf{1.81} \pm 0.14$&$1.09 \pm 0.06$&$\mathbf{44.90} \pm 0.90$&$\mathbf{90.46\%} \pm 0.95\%$\\
			\hline\hline	
	\end{tabular}}
	\label{tab:TMD&BVTV360}
\end{table*}
\begin{table*}[ht!]
	\caption{Statistics of LR-QCT data: mean $\pm$ half range of test sets of 5-fold cross-validation. In bold (except for SD): best values per setting.}
	\centering
	\resizebox{\linewidth}{!}
	{\begin{tabular}{ll|ccccccc}
			Parameter&Filter&$\Delta$ Avg &SD &RMSE&AE&PE&PSNR&adj.R$^2$\\ 
			\hline\hline
			\multirow{4}{60pt}{per voxel [$\mgcc$]}
			&$-$&$1.47 \pm 0.83$&$332.26 \pm 0.86$&$338.67 \pm 1.14$&$\le87.45 \pm 0.59$&$123.68 \pm 0.84$&$2.36 \pm 0.17$&$1.97 \% \pm 0.15\%$\\
			&BM4D&$1.41 \pm 0.89$&$314.44 \pm 0.90$&$321.42 \pm 1.16$&$\mathbf{\le87.21} \pm 0.58$&$123.33 \pm 0.82$&$3.41 \pm 0.18$&$2.15 \% \pm 0.16\%$\\
			&$\NNBMD$&$0.14 \pm 1.76$&$100.47 \pm 1.53$&$\mathbf{141.99} \pm 1.10$&$104.43 \pm 1.28$&$\mathbf{112.61} \pm 0.94$&$\mathbf{19.75} \pm 0.17$&$\mathbf{5.67\%} \pm 0.30\%$\\
			&$\NNSP$&$\mathbf{-0.82} \pm 1.28$&$129.65 \pm 0.98$&$165.00 \pm 1.19$&$100.22 \pm 1.61$&$118.91 \pm 1.41$&$16.75 \pm 0.05$&$3.03 \% \pm 0.33\%$\\
			
			\hdashline
			\multirow{4}{50pt}{BMD [$\mgcc$]}
			&$-$&$2.07 \pm 0.37$&$34.29 \pm 0.57$&$11.48 \pm 0.59$&$\mathbf{5.76} \pm 0.59$&$9.19 \pm 0.37$&$43.91 \pm 0.97$&$89.46 \% \pm 1.01\%$\\
			&BM4D&$2.00 \pm 0.37$&$34.26 \pm 0.59$&$11.38 \pm 0.59$&$\mathbf{5.76} \pm 0.58$&$9.10 \pm 0.37$&$44.07 \pm 0.98$&$89.60 \% \pm 1.00\%$\\
			&$\NNBMD$&$\mathbf{0.30} \pm 1.21$&$32.19 \pm 0.81$&$\mathbf{8.36} \pm 0.38$&$6.59 \pm 0.40$&$\mathbf{5.08} \pm 0.14$&$\mathbf{50.23} \pm 0.68$&$\mathbf{94.26\%} \pm 0.54\%$\\
			&$\NNSP$&$-0.44 \pm 1.35$&$32.87 \pm 0.96$&$8.48 \pm 0.41$&$6.51 \pm 0.38$&$5.26 \pm 0.16$&$49.95 \pm 0.96$&$94.00 \% \pm 0.49\%$\\
			
			\hdashline
			\multirow{4}{50pt}{TMD [$\mgcc$]}
			&$-$&$119.08 \pm 0.72$&$30.53 \pm 0.64$&$125.54 \pm 0.92$&$124.66 \pm 0.97$&$\mathbf{13.42} \pm 0.25$&$-6.77 \pm 0.57$&$1.73 \% \pm 0.87\%$\\
			&BM4D&$104.97 \pm 0.70$&$29.03 \pm 0.67$&$111.79 \pm 0.89$&$110.84 \pm 0.94$&$13.84 \pm 0.26$&$-4.45 \pm 0.57$&$2.12 \% \pm 0.91\%$\\
			&$\NNBMD$&$-43.68 \pm 5.51$&$15.49 \pm 3.18$&$49.99 \pm 4.75$&$49.39 \pm 4.89$&$13.93 \pm 2.63$&$11.68 \pm 1.40$&$34.28 \% \pm 5.02\%$\\
			&$\NNSP$&$\mathbf{0.56} \pm 2.49$&$22.97 \pm 0.92$&$\mathbf{21.53} \pm 0.86$&$\mathbf{17.68} \pm 0.89$&$14.76 \pm 0.61$&$\mathbf{28.50} \pm 1.43$&$\mathbf{48.67\%} \pm 6.23\%$\\
			
			\hdashline
			\multirow{4}{50pt}{BV/TV [$\%$]}
			&$-$&$19.99 \pm 0.08$&$4.15 \pm 0.09$&$20.42 \pm 0.09$&$20.36 \pm 0.09$&$2.36 \pm 0.05$&$-0.07 \pm 0.69$&$66.12 \% \pm 2.00\%$\\
			&BM4D&$19.28 \pm 0.09$&$4.36 \pm 0.09$&$19.70 \pm 0.10$&$19.64 \pm 0.10$&$2.35 \pm 0.05$&$0.64 \pm 0.68$&$66.83 \% \pm 2.04\%$\\
			&$\NNBMD$&$-4.18 \pm 0.62$&$7.54 \pm 0.47$&$5.03 \pm 0.45$&$4.79 \pm 0.47$&$\mathbf{1.27} \pm 0.05$&$28.02 \pm 2.29$&$86.66 \% \pm 1.24\%$\\
			&$\NNSP$&$\mathbf{-0.02} \pm 0.43$&$6.38 \pm 0.24$&$\mathbf{2.48} \pm 0.14$&$\mathbf{2.04} \pm 0.12$&$1.38 \pm 0.09$&$\mathbf{42.14} \pm 0.92$&$\mathbf{86.86\%} \pm 1.20\%$\\
			\hline\hline	
	\end{tabular}}
	\label{tab:TMD&BVTV100}
\end{table*}
\begin{table}
	\caption{Avg (SD) from HRpQCT of the 2274 analyzed patches, TMD and BV/TV are computed with $t=225~\mgcc$.}
	\centering
	\resizebox{\linewidth}{!}
	{\begin{tabular}{cccc}
			per voxel[$\mgcc$] & BMD[$\mgcc$] & TMD[$\mgcc$] & BV/TV[$\%$]\\\hline\\[-2ex]
			$115.90\;(118.57)$ & $115.90\;(35.14)$ & $334.54\;(29.73)$ &$17.30\;(6.93)$
	\end{tabular}}
	\label{tab:TMD&BVTVxct}
\end{table}
Means and half-ranges of the 5-fold cross-validation analysis conducted over the transformations of the HR- and LR-QCT data are shown in table~\ref{tab:TMD&BVTV360} and~\ref{tab:TMD&BVTV100} and reference HRpQCT values are reported in table~\ref{tab:TMD&BVTVxct} (by definition the average per voxel is the same as for BMDs).
For the statistics per voxel and BMD, no specific filter showed definite improvement.
However when considering only RMSE, PE and PSNR and adj.~R$^2$, best values were obtained with $\NNBMD$ (and nearly as good with $\NNSP$ on LR-QCT).
In particular BMD statistics computed on the filtered LR-QCT data were close to those computed on raw HR-QCT. 
On the other hand, accuracy errors were generally lower on the un-filtered or BM4D-filtered data, than after filtering with the proposed neural networks, specifically for the per-voxel statistics.

Statistics of TMD and BV/TV were more evident:
We obtained notably lower errors with $\NNSP$ from LR-QCT than from HR-QCT without filtering.  
$\NNSP$ reduced the RMSEs of TMD by factor $39\%$ (HR-QCT) and $83\%$ (LR-QCT) and the RMSE of the filtered LR-QCT was still by $25\%$ reduced in comparison to the RMSE of raw HR-QCT.
PSNRs of TMD increased by $9.76$ (HR-QCT) and $35.27~\db$ (LR-QCT) with a difference of $5.84~\db$ between filtered LR-QCT and raw HR-QCT.
$\NNBMD$ on the other hand improved statistics of TMD only on LR-QCT.
Similarly, both neural networks improved estimation of BV/TV with lowest RMSEs and AEs and highest PSNRs on $\NNSP$.
Precision of TMD did not clearly improve nor decline for any of the considered filters but both neural networks decreased precision errors of BV/TV.

\begin{figure*}
	\centering
	\includegraphics[width=\linewidth]{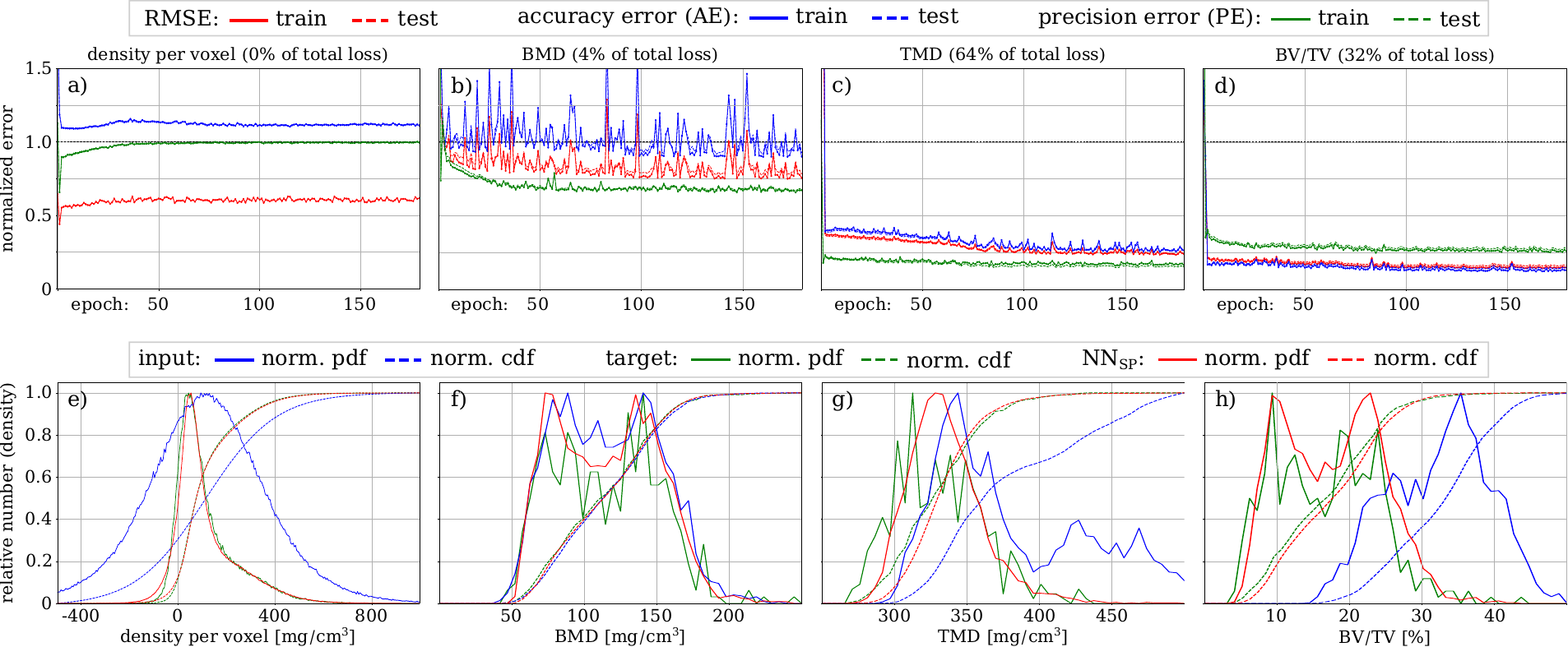}
	\caption{Error metrics and histograms for statistics per voxel, BMD, TMD and BV/TV for $\NNSP$. (a-d): RMSE, AE and PE on training and test set. (e-h): Normalized (norm. pdf) and cumulative histogram (norm. cdf) of input, target and output samples.} 
	\label{fig:LearningCurves}
\end{figure*}
Figure~\ref{fig:LearningCurves}~(a-d) shows normalized error metrics of the training and test set of each epoch for $t=225~\mgcc$, averaged over all nine repetitions of LR-, MR- and HR-QCT.
Errors per voxel~(a) and per BMD~(b) have been not or only slightly improved when comparing to initial errors,  
but TMD (c) and BV/TV (d), for which $\NNSP$ was mainly designed, show both a noticeable error reduction already on early epochs.
Differences between errors (c-d) computed on the training- and test-test were small or nearly undetectable. 
We observed even cases with smaller errors on the test than on the training set (for instance PE on TMD), showing that no over-fitting took place but optimum capacity to generalize instead. 

Figure~\ref{fig:LearningCurves}~(e-h) shows normalized histograms, a visual representation of the global distribution of each parameter map.
All histograms of the filtered data coincide mostly with those of the target data. 
In particular the histogram of the direct output~(e) transforms the rather Gaussian-like histogram of the input data to one that contains the characteristic right tail of HRpQCT resolution.
Noteworthy, BMD~(f) has already been computed with high precision and accuracy on the raw data, explaining the small error reduction with $\NNSP$~(b).
TMD and BV/TV contained, in contrast to the voxel-wise metrics and BMD, a tube-current-dependent offset in the raw data which was compensated by the neural network.

\section{Discussion and Conclusion}
\begin{figure}
	\centering
	\includegraphics[width=\linewidth]{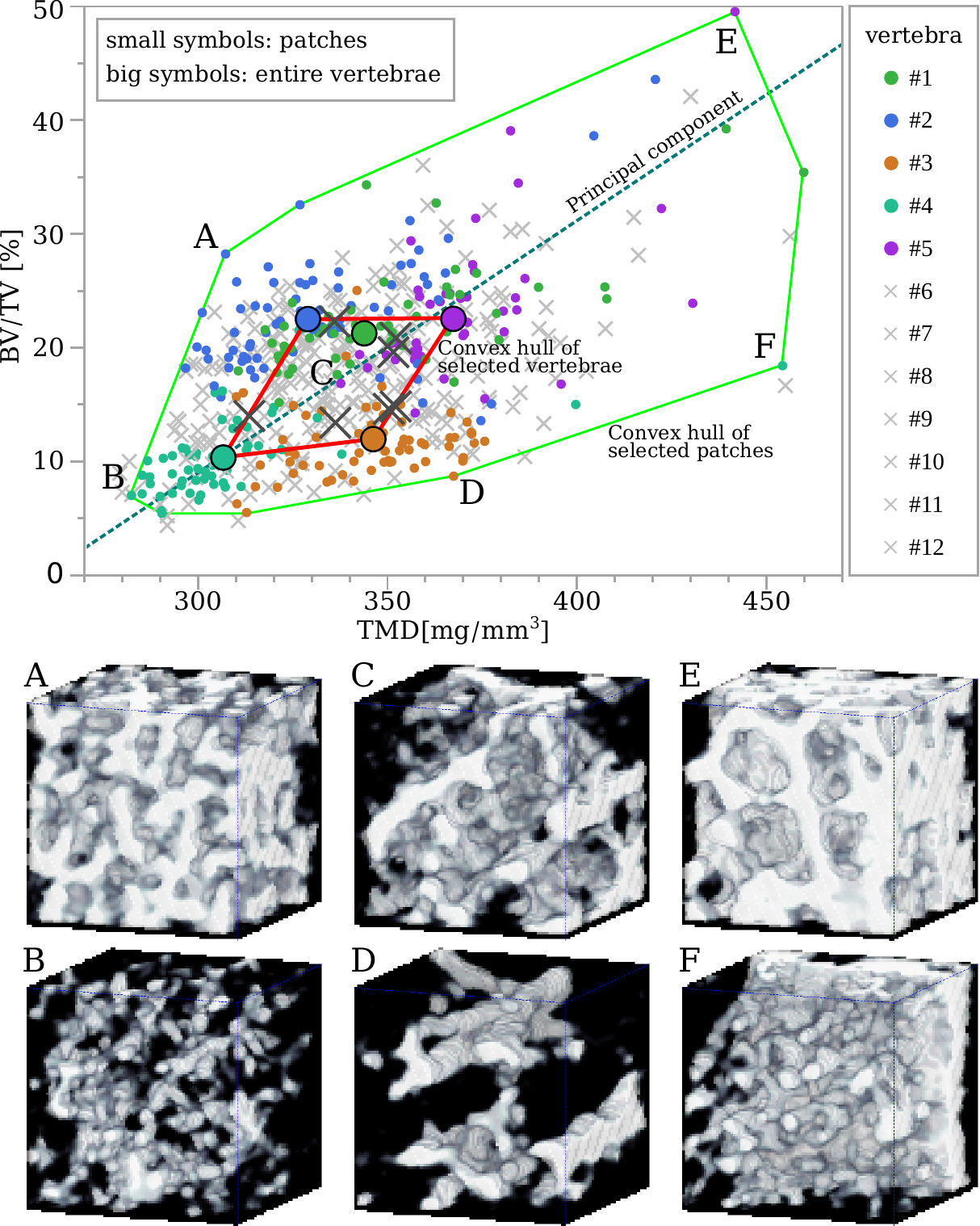}
	\caption{TMD vs. BV/TV with $t=225~\mgcc$: The convex hull of the five selected vertebrae (red line) includes all 12 vertebrae, the convex hull of the patches from the five vertebrae (green line) includes $99.5\%$ of all patches.
		Bottom: Selected samples.}
	\label{fig:distPatches}
\end{figure}
We demonstrated the feasibility to compute TMD and BV/TV with high accuracy and precision on high- and low-resolution QCT.
Accurate computation of TMD can be considered impossible on unfiltered low or mid-resolution QCT, and maybe still on HR-QCT
(adj. R$^2$: $1.73$ to $15.90\%$).
After filtering with the presented method, we obtained adjusted coefficients of determination of $48.67\%$ to $65.30\%$, that means filtered low-resolution QCTs revealed more information than unfiltered high resolution QCT, though is still affected with errors when comparing to HRpQCT.
A similar error reduction has been observed for BV/TV over the entire range of conducted statistics.
We obtained an error reduction of up to $83\%$ for TMD and $88\%$ for BV/TV when applied on low-resolution QCT, which is an important improvement for those parameters on low-resolution QCT.
Two denoising filters of comparison, the BM4D filter and a neural network, performed for TMD only similar or even worse than the unfiltered volume.
We showed that TMD and BV/TV require, at least for LR-QCT, a specific and dedicated goal-driven filter since general denoising methods or unfiltered volumes are only of limited use here.

Noise reduction or recovery of fine scale information is only possible in a limited range of signal-to-noise ratios.
Decades of research did eventually find near optimum solutions of general noise reduction of images with standard noise levels. 
A comparison of state-of-the art image noise reduction filters~\cite{Chen2017Low2} on images with PSNRs around $30~\db$, including BM3D and three neural networks, showed $3$ to $23\%$ ($0.2$ to $2.25~\db$) better performance on neural networks than filtering with BM3D.
The volumes we used in this study contained low PSNRs between $2.36$ to $12.95~\db$ and BM4D gained only small improvements here.
By using $\NNBMD$ we obtained an increase of $8.5$ to $17.39~\db$ (PSNR per voxel).
Those numbers demonstrate the strength of neural networks in particular for \emph{in vivo} CT imaging which is due to the correlation of ray exposure and image quality prone to low SNRs.  

We conducted a-priori some tests to design the reported architecture of the neural network and to find its used hyper-parameters.
These parameters were then fixed for all consecutive analyses of the final network.   
1) We selected the reported architecture after evaluating designs based on convolution masks between $3\times3\times1$ and $9\times 9 \times 5$ voxels and networks with equal, increasing or decreasing numbers of convolutions per layer and those with a decreased or increased number in the center layer. 
2) We tested as well loss functions based one the precision and accuracy errors or the structural similarity index~\cite{You2018structurally}, also the Kullback-Leibner distance could be used.
Since performance did not improve significantly we used the more common MSE-loss.
3) The reported learning rate and batch size per weight adjustment have been derived after applying a two dimensional grid search as those least prone for over-fitting or -generalizing.

Additionally to $\NNBMD$ and $\NNSP$ we analyzed neural networks trained only with one loss of BMD, TMD or BV/TV, we obtained in particular on the network trained with the TMD-loss higher errors on TMD than with $\NNSP$, that means that the additional BV/TV-loss used in $\NNSP$ served as a regularization function that influenced not only errors of BV/TV but also of TMD. 
We analyzed then neural networks not to perform denoising but to immediately compute maps of BMD, TMD and BV/TV, those showed to be prone for over-fitting.
Neural networks for the direct computation of structural maps are neighborhood ($d$) and threshold ($t$) specific, since they implement an instance of a certain local structural parameter (e.g. $\TMD_{d,t}(V)$, see Eq.~\ref{Eq:locTMD}). 
In contrast, these parameters were multiple times involved when training $\NNSP$ (multi-threshold and/or multi-neighborhood), but only once  when computing TMD directly. 
Thus the application of multiple instances of loss functions is likely an important constraint of regularization that applies only for neural networks of denoising.

The data set of this prototype study was designed 1) to obtain a sufficiently large set of patches with a realistic variance (structural information) 2) to study the interaction between noise and bone structure using repetitions of each phantom and 3) to analyze the influence of noise levels from different tube currents. 
The vertebra phantoms were considered as a pile of textural patches serving for local structural analyses.
In order to reduce the work-load of the study and to exclude unnecessary error-sources, we reduced the number of vertebrae phantoms from twelve to five prototype phantoms, thereby reducing the number of scans from $120$ to $50$ but maintaining the structural information of the entire set of phantoms:  
Fig.~\ref{fig:distPatches} shows the structural variance of sample patches, and the distribution BV/TV vs. TMD of a sparse subset of all twelve phantoms that were initially scanned on the HR-pQCT device.
When considering parameters of the entire vertebrae (big symbols), all twelve could be interpolated from the five selected ones (colored symbols).
Prototype patches (small colored symbols) occupied each a specific area in parameter space but their variance per-vertebra was high enough to fill gaps between vertebrae.
Only $0.5\%$ of all patches were outside the convex hull of prototypes (green line).
In conclusion, adding further prototype phantoms would not allow to assess more patch-wise structural information but instead unnecessarily increase the complexity and error-proneness of the experiment.   
However, further experiments that allowed to leave out entire vertebrae phantoms for testing might be sensible before applying the method in medical practice.
Since typically twenty or more vertebrae are used for such a per-vertebra analysis, the entire experiment required a massive effort when including additionally repetitions, variations of tube currents and different CT scanners.
Alternatively, generative methods~\cite{Iarussi2020GAN} could soon offer a virtually infinite number of realistic in-silico patches.
These could be used to enlarge the training set and leaving the costly real scans for fine-tuning and testing only.

\section*{Acknowledgment}
We thank C.-C.~Gl\"uer for providing the phantoms and N.~Abdullayev for assisting in the scans.
This work was supported by Agencia Nacional de Promoci\'on Cient\'ifica y Tecnol\'ogica (ANPCyT) PICT 2017-1731. 

\section*{References}

\bibliography{Bibshort}

\end{document}